\RequirePackage[dvipsnames]{xcolor}
\documentclass[a4paper]{easychair}

\usepackage{enumitem}
\setitemize{topsep=0em,noitemsep}

\usepackage[utf8]{inputenc}

\usepackage{mycomments}
\usepackage{mybiblio}
\usepackage{myhyperref}
\usepackage{myrefs}
\usepackage{mylistings}
\usepackage{mymath}

\usepackage{notations}

\title{Tail Modulo Cons}

\author{
Fr{\'e}d{\'e}ric Bour\inst{1}\inst{2}
\and
Basile Cl{\'e}ment\inst{1}
\and
Gabriel Scherer\inst{1}
}

\institute{
INRIA
\and
Tarides
}

\authorrunning{Bour, Cl{\'ement}, Scherer}
\titlerunning{Tail Modulo Cons}

\begin{document}

\maketitle

\begin{abstract}
  OCaml function calls consume space on the system stack. Operating
  systems set default limits on the stack space which are much lower
  than the available memory. If a program runs out of stack space,
  they get the dreaded ``Stack Overflow'' exception -- they crash. As
  a result, OCaml programmers have to be careful, when they write
  recursive functions, to remain in the so-called
  \emph{tail-recursive} fragment, using \emph{tail} calls that do not
  consume stack space.

  This discipline is a source of difficulties for both beginners and
  experts. Beginners have to be taught recursion, and then
  tail-recursion. Experts disagree on the ``right'' way to write
  \code{List.map}. The direct version is beautiful but not
  tail-recursive, so it crashes on larger inputs. The naive
  tail-recursive transformation is (slightly) slower than the direct
  version, and experts may want to avoid that cost. Some libraries
  propose horrible implementations, unrolling code by hand, to
  compensate for this performance loss. In general, tail-recursion
  requires the programmer to manually perform sophisticated program
  transformations.

  In this work we propose an implementation of ``Tail Modulo Cons''
  (TMC) for OCaml. TMC is a program transformation for a fragment of
  non-tail-recursive functions, that rewrites them in
  \emph{destination-passing style}. The supported fragment is smaller
  than other approaches such as continuation-passing-style, but the
  performance of the transformed code is on par with the direct,
  non-tail-recursive version. Many useful functions that traverse
  a recursive datastructure and rebuild another recursive structure
  are in the TMC fragment, in particular \code{List.map}
  (and \code{List.\{filter,append\}}, etc.). Finally those functions
  can be written in a way that is beautiful, correct on all inputs,
  and efficient.

  In this work we give a novel modular, compositional definition of
  the TMC transformation. We discuss the design space of
  user-interface choices: what degree of control for the user, when to
  warn or fail when the transformation may lead unexpected results. We
  mention a few remaining design difficulties, and present
  (in appendices) a performance evaluation of the transformed code.
\end{abstract}

\section{Introduction}

\subsection{Prologue}

``OCaml'', we teach our students, ``is a functional programming language. We can write the beautiful function \code{List.map} as follows:''
\begin{lstlisting}
let rec map f = function
| [] -> []
| x :: xs -> f x :: map f xs
\end{lstlisting}

``Well, actually'', we continue, ``OCaml is an effectful language, so
we need to be careful about the evaluation order. We want \code{map}
to process elements from the beginning to the end of the input list,
and the evaluation order of \lstinline{f x :: map f xs} is
unspecified. So we write:
\pagebreak
\begin{lstlisting}
let rec map f = function
| [] -> []
| x :: xs ->
  let y = f x in
  y :: map f xs
\end{lstlisting}

``Well, actually, this version fails with a \code{Stack_overflow}
exception on large input lists. If you want your \code{map} to behave
correctly on all inputs, you should write a \emph{tail-recursive}
version. For this you can use the accumulator-passing style:''
\begin{lstlisting}
let map f li =
  let rec map_ acc = function
  | [] -> List.rev acc
  | x :: xs -> map_ (f x :: acc) xs
  in map_ [] f li
\end{lstlisting}

``Well, actually, this version works fine on large lists, but it is
less efficient than the original version. It is noticeably slower on
small lists, which are the most common inputs for most programs. We
measured it 35\% slower on lists of size 10. If you want to write
a robust function for a standard library, you may want to support both
use-cases as well as possible. One approach is to start with
a non-tail-recursive version, and switch to a tail-recursive version
for large inputs; even there you can use some manual optimizations to
reduce the overhead of the accumulator. For example, the nice
\href{https://github.com/c-cube/ocaml-containers}{Containers} library
does it as follows:''.

\lstset{basicstyle=\tiny\ttfamily}
\begin{minipage}{0.6\linewidth}
\begin{lstlisting}
let tail_map f l =
  (* Unwind the list of tuples, reconstructing the full list front-to-back.
     @param tail_acc a suffix of the final list; we append tuples' content
     at the front of it *)
  let rec rebuild tail_acc = function
    | [] -> tail_acc
    | (y0, y1, y2, y3, y4, y5, y6, y7, y8) :: bs ->
      rebuild (y0 :: y1 :: y2 :: y3 :: y4 :: y5 :: y6 :: y7 :: y8 :: tail_acc) bs
  in
  (* Create a compressed reverse-list representation using tuples
     @param tuple_acc a reverse list of chunks mapped with [f] *)
  let rec dive tuple_acc = function
    | x0 :: x1 :: x2 :: x3 :: x4 :: x5 :: x6 :: x7 :: x8 :: xs ->
      let y0 = f x0 in let y1 = f x1 in let y2 = f x2 in
      let y3 = f x3 in let y4 = f x4 in let y5 = f x5 in
      let y6 = f x6 in let y7 = f x7 in let y8 = f x8 in
      dive ((y0, y1, y2, y3, y4, y5, y6, y7, y8) :: tuple_acc) xs
    | xs ->
      (* Reverse direction, finishing off with a direct map *)
      let tail = List.map f xs in
      rebuild tail tuple_acc
  in
  dive [] l
\end{lstlisting}
\end{minipage}
\hfill
\begin{minipage}{0.4\linewidth}
\begin{lstlisting}
let direct_depth_default_ = 1000

let map f l =
  let rec direct f i l = match l with
    | [] -> []
    | [x] -> [f x]
    | [x1;x2] -> let y1 = f x1 in [y1; f x2]
    | [x1;x2;x3] ->
      let y1 = f x1 in let y2 = f x2 in [y1; y2; f x3]
    | _ when i=0 -> tail_map f l
    | x1::x2::x3::x4::l' ->
      let y1 = f x1 in
      let y2 = f x2 in
      let y3 = f x3 in
      let y4 = f x4 in
      y1 :: y2 :: y3 :: y4 :: direct f (i-1) l'
  in
  direct f direct_depth_default_ l
\end{lstlisting}
\end{minipage}
\lstset{basicstyle=\small\ttfamily}

At this point, unfortunately, some students leave the class and never
come back. (These days they just have to disconnect from the
remote-teaching server.)

We propose a new feature for the OCaml compiler, an explicit, opt-in
``Tail Modulo Cons'' transformation, to retain our students. After the
first version (or maybe, if we are teaching an advanced class, after
the second version), we could show them the following version:
\begin{lstlisting}
let[@tail_mod_cons] rec map f = function
| [] -> []
| x :: xs -> f x :: map f xs
\end{lstlisting}

This version would be as fast as the simple implementation,
tail-recursive, and easy to write.

The catch, of course, is to teach when this \code{[@tail_mod_cons]}
annotation can be used. Maybe we would not show it at all, and pretend
that the direct \code{map} version with \code{let y} is fine. This
would be a much smaller lie than it currently is,
a \code{[@tail_mod_cons]}-sized lie.

Finally, experts would be very happy. They know about all these
versions, but they would not have to write them by hand anymore. Have
a program perform (some of) the program transformations that they are
currently doing manually.

\subsection{TMC transformation example}

A function call is in \emph{tail position} within a function
definition if the definition has ``nothing to do'' after evaluating
the function call -- the result of the call is the result of the whole
function at this point of the program. (A precise definition will be
given in Section~\ref{sec:tcmc}.) A function is \emph{tail recursive}
if all its recursive calls are tail calls.

In the definition of \code{map}, the recursive call is not in tail
position: after computing the result of \code{map f xs} we still have
to compute the final list cell, \code{y ::$\ \hole$}. We say that a call is
\emph{tail modulo cons} when the work remaining is formed of data
\emph{constructors} only, such as \code{(::)} here.

\begin{lstlisting}
let[@tail_mod_cons] rec map f = function
| [] -> []
| x :: xs ->
  let y = f x in
  y :: map f xs
\end{lstlisting}

Other datatype constructors may also be used; the following example is
also tail-recursive \emph{modulo cons}:

\begin{lstlisting}
let[@tail_mod_cons] rec tree_of_list = function
| [] -> Empty
| x :: xs -> Node(Empty, x, tree_of_list xs)
\end{lstlisting}

The TMC transformation returns an equivalent function in
\emph{destination-passing} style where the calls in \emph{tail modulo
  cons} position have been turned into \emph{tail} calls. In
particular, for \code{map} it gives a tail-recursive function, which
runs in constant stack space; many other list functions also become
tail-recursive. The transformed code of \code{map} can be described as
follows:

\hspace{-1.6em}
\begin{minipage}{0.5\linewidth}
\begin{lstlisting}
let rec map f = function
| [] -> []
| x::xs ->
  let y = f x in
  let dst = y :: Hole in
  map_dps dst 1 f xs;
  dst
\end{lstlisting}
\end{minipage}
\hfill
\begin{minipage}{0.5\linewidth}
\begin{lstlisting}
and map_dps dst i f = function
| [] ->
  dst.i <- []
| x::xs ->
  let y = f x in
  let dst' = y :: Hole in
  dst.i <- dst';
  map_dps dst' 1 f xs
\end{lstlisting}
\end{minipage}

The transformed code has two variants of the \code{map} function. The
\code{map_dps} variant is in \emph{destination-passing style}, it
expects additional parameters that specify a memory location,
a \emph{destination}, and will write its result to this
\emph{destination} instead of returning it. It is tail-recursive. The
\code{map} variant provides the same interface as the non-transformed
function, and internally calls \code{map_dps} on non-empty lists. It
is not tail-recursive, but it does not call itself recursively, it
jumps to the tail-recursive \code{map_dps} after one call.

The key idea of the transformation is that the expression \code{y ::
  map f xs}, which contained a non-tail-recursive call, is transformed
into first the computation of a \emph{partial} list cell, written %
\code{y :: Hole}, followed by a call to \code{map_dps} that is asked
to write its result in the position of the \code{Hole}. The recursive
call thus happens after the cell creation (instead of before), in
tail-recursive position in the \code{map_dps} variant. In the direct
variant, the value of the destination \code{dst} has to be returned
after the call.

The transformed code is in a pseudo-OCaml, it is not a valid OCaml
program: we use a magical \code{Hole} constant, and our notation
\code{dst.i <- ...} to update constructor parameters in-place is also
invalid in source programs. The transformation is implemented on
a lower-level, untyped intermediate representation of the OCaml
compiler (Lambda), where those operations do exist. The OCaml type
system is not expressive enough to type-check the transformed program:
the list cell is only partially-initialized at first, each partial
cell is mutated exactly once, and in the end the whole result is
returned as an \emph{immutable} list. Some type system are expressive
enough to represent this transformed code, notably Mezzo
\citep*{mezzo}.

\subsection{Other approaches}

\subsubsection{More general transformations}

Instead of a program transformation in \emph{destination-passing}
style, we could perform a more general program transformation that can
make more functions tail-recursive, for example a generic
\emph{continuation-passing} style (CPS) transformation. We have three
arguments for implementing the TMC transformation:
\begin{itemize}
\item The TMC transformation generates more efficient code, using
  mutation instead of function calls. On the OCaml runtime, the
  difference is a large constant factor.\footnote{On a toy benchmark with
  large-sized lists, the CPS version is 100\% slower and has 130\%
  more allocations than the non-tail-recursive version.}

\item The CPS transformation can be expressed at the source level, and
  can be made reasonably nice-looking using some monadic-binding
  syntactic sugar. TMC can only be done by the compiler, or using
  safety-breaking features.

\item TMC is provided as an opt-in, on-demand optimization. We can add
  more such optimizations, they are not competing with each other,
  especially if they are to be rather used by expert
  programmers. Someone should try presenting CPS as an
  annotation-driven transformation, but we wanted to look at TMC
  first.
\end{itemize}

\subsubsection{Different runtimes}

Using the native system stack is a choice of the OCaml
implementation. Some other implementations of functional languages,
such as SML/NJ, use a different stack (the OCaml bytecode interpreter
also does this), or directly allocate stack frames on their GC-managed
heap. This approach makes ``stack overflow'' go away completely, and
it also makes it very simple to implement stack-capture control
operators, such as continuations, or other stack operations such as
continuation marks.

On the other hand, using the native stack brings compatibility
benefits (coherent stack traces for mixed OCaml+C programs), and seems
to noticeably improve the performance of function calls (on benchmarks
that are only testing function calls and return, such as Ackermann or
the naive Fibonacci, OCaml can be 4x, 5x faster than SML/NJ.)

Lazy (call-by-need) languages will also often avoid running into stack
overflows: as soon as a lazy datastructure is returned, which is the
default, functions such as \code{map} will return immediately, with
recursive calls frozen in a lazy thunk, waiting to be evaluated
on-demand as the user traverses the result structure. User still need
to worry about tail-recursivity for their strict functions (if the
implementation uses the system stack); strict functions are often
preferred when writing performant code.

\subsubsection{Unlimiting the stack}

Some operating systems can provide an unlimited system stack; such as
\code{ulimit -s unlimited} on Linux systems -- the system stack is
then resized on-demand. Then it is possible to run non-tail-recursive
functions without fear of overflows. Frustratingly, unlimited stacks
are not available on all systems, and not the default on any system in
wide use. Convincing all users to setup their system in a non-standard
way would be \emph{much} harder than performing a program
transformation or accepting the CPS overhead for some
programs.

\subsection{Related Work}

Tail-recursion modulo cons was well-known in the Lisp community as
early as the 1970s. For example the REMREC system~\citep*{remrec}
would automatically transform recursive functions into loops, and
supports modulo-cons tail recursion. It also supports tail-recursion
modulo associative arithmetic operators, which is outside the scope
of our work, but supported by the GCC compiler for example. The TMC
fragment is precisely described (in prose) in \citet*{unwinding}.

In the Prolog community it is a common pattern to implement
destination-passing style through unification variables; in particular
``difference lists'' are a common representation of lists with a final
hole. Unification variables are first-class values, in particular they
can be passed as function arguments. This makes it easy to write the
destination-passing-style equivalent of a context of the form
\code{List.append li $\ \hole$}, as the difference list
\code{(List.append li X, X)}. In constrast, we only support direct
constructor applications. However, this expressivity comes at
a performance cost, and there is no static checking that the data is
fully initialized at the end of computation.

In general, if we think of non-tail recursive functions as having an
``evaluation context'' left for after the recursive call, then the
techniques to turn classes of calls into tail-calls correspond to
different reified representations of non-tail contexts, as long as
they support efficient composition and hole-plugging. TMC comes from
representing data-construction contexts as the partial data itself,
with hole-plugging by mutation. Associative-operator transformations
represent the context $1 + (4 + \hole)$ as the number $5$
directly. (Sometimes it suffices to keep around an abstraction of the
context; this is a key idea in John Clements' work on stack-based
security in presence of tail calls.)

\cite*{minamide} gives a ``functional'' interface to
destination-passing-style program, by presenting a partial
data-constructor composition \code{Foo(x,Bar($\hole$))} as a use-once,
linear-typed function \code{linfun h -> Foo(x,Bar(h))}. Those special
linear functions remain implemented as partial data, but they
expose a referentially-transparent interface to the programmer,
restricted by a linear type discline. This is a beautiful way to
represent destination-passing style, orthogonal to our work: users of
Minamide's system would still have to write the transformed version by
hand, and we could implement a transformation into destination-passing
style expressed in his system. \cite*{mezzo} supports a more
general-purpose type system based on separation logic, which can
directly express uniquely-owned partially-initialized data, and its
implicit transformation into immutable, duplicable results. (See the
\href{https://protz.github.io/mezzo/code_samples/list.mz.html}{List}
module of the Mezzo standard library, and in particular \code{cell},
\code{freeze} and \code{append} in destination-passing-style).

\subsection{Contributions}

This work is in progress. We claim the following contributions:
\begin{itemize}
\item A formal grammar of which programs expressions are in
  the ``Tail Modulo Cons'' fragment.

\item A novel, modular definition of the transformation
  into destination-passing-style.

\item Discussion of the user-interface issues related to
  transformation control.

\item A performance evaluation of the transformation for
  \code{List.map}, in the specific context of the OCaml runtime.
\end{itemize}

A notable non-contribution is a correctness proof for the
transformation. We would like to work on a correctness proof soon; the
correctness argument requires reasoning on mutability and ownership of
partial values, an excellent use-case for separation logic.

\section{Tail Calls Modulo Constructors}
\label{sec:tcmc}

\begin{mathparfig}{fig:langgram}{A first-order programming language}
  \begin{array}{r@{~}r@{~}l}
    \Set{Exprs} \ni e, d
    & \bnfeq
    & x, y
    \\
    & \bnfor
    & n \in \mathbb{N}
    \\
    & \bnfor
    & \app f e
    \\
    & \bnfor
    & \letin x e {e'}
    \\
    & \bnfor
    & \constr K {\fam i {e_i}}
    \\
    & \bnfor
    & \match e {\fam i {\clause {p_i} {e'_i}}}
    \\
    & \bnfor
    & \setref d e {e'}
  \end{array}

  \begin{array}{r@{~}r@{~}l}
    \Set{FunctionNames} \ni f
    \\
    \Set{Patterns} \ni p
    & \bnfeq
    & x \bnfor \constr K {\fam i {p_i}}
	\end{array}

	\begin{array}{r@{~}r@{~}l}
		\Set{Stmt} \ni s
		& \bnfeq
		& {\letrecfam i {f_i} x {e_i}}
  \end{array}
\end{mathparfig}

In order to simplify the presentation of the transformation, we
consider a simple untyped language, described in
Figure~\ref{fig:langgram}.  This language, which we present with
a syntax similar to OCaml, embeds function application and sequencing
let-binding, as well as constructor application and pattern-matching.
In addition, to implement the imperative DPS transformation, we
include a special operator:  $\setref d e {e'}$ is an imperative construct which updates
$d$'s $e$-th argument in-place.  All those constructs (and more) are
present in the untyped intermediate language used in the OCaml
compiler where we implemented the transformation.  One notable missing
construct is function abstraction. In fact, this model requires that
all functions be called by name, and functions can only be defined
through a toplevel $\kwd{let}~\kwd{rec}$ statement. Our implementation
supports the full OCaml language, but it cannot specialize
higher-order function arguments for TMC.

In the following, we will use syntaxic sugar for some usual
constructs; namely, we can desugar $\seq {e1} {e2}$ into the let-binding
$\letin \wild {e1} {e2}$ (where $\wild$ is a fresh variable name) and
$(e_1, \dots, e_n)$ into the constructor application
$\constr {\kwd{Tuple}} {(e_1, \dots, e_n)}$.


The multi-hole grammar of tail contexts, where each hole indicates a
tail position, for this simple language is depicted in
Figure~\ref{fig:tailgram}.  To interpret a multi-hole context $\Tcont$
with $n$ holes, we denote by $\plug \Tcont {e_1, \dots, e_n}$ the term
obtained by replacing each of the holes in $\Tcont$ from left to right
with the expressions $e_1, \dots, e_n$.  In a decomposition $e = \plug
\Tcont {e_1, \dots, e_n}$, each of the $e_i$ is in tail position; in
particular, a call $\app f e$ is in tail position (i.e. it is a tail call) in
expression $e'$ if there is a decomposition of $e'$ as $\plug \Tcont
{e_1, \dots, e_j, \app f e, e_{j + 1}, \dots, e_n}$.

One can remark that, for a language construct, the holes in the tail
context are precisely the complement of holes in the evaluation
context.  For instance, the construct $\letin x e {e'}$ has evaluation
contexts of the form $\letin x E {e'}$ and tail contexts
$\letin x e T$.  In some sense, the tail contexts are ``guarded'' by
the evaluation context: a reduction can only occur in a tail position
after the construct has been reduced away, and the subterm in tail
position is now at toplevel. This guarantees that when we start
reducing inside the tail context, there is no remaining computation to
be performed in the surrounding context. The ``depth'' of the
surrounding context is a source-level notion that is directly related
to call-stack size: tail calls do not require reserving frames on the
call stack.

\begin{mathparfig}{fig:tailgram}{Tail multicontexts, constructor contexts}
  \label{fig:consgram}
  \begin{array}{r@{~}r@{~}l@{\quad}l}
    \Set{TailCtx} \ni \Tcont
    & \bnfeq
    &
    \\
    & \bnfor
    & \hole
    &
    \\
    & \bnfor
    & \seq e \Tcont
    &
    \\
    & \bnfor
    & \letin x e \Tcont
    &
    \\
    & \bnfor
    & \match e {\fam j {\clause {p_j} \Tcont_j}}
    &
    \\
    \\
    \Set{ConstrCtx} \ni {\plug \Ccont \hole}
    & \bnfeq
    & \hole
    \\
    & \bnfor
    & \constr K {(\fam i {e_i}, \plug \Ccont \hole, \fam j {e_j})}
  \end{array}
\end{mathparfig}

Our first goal is to figure out what the proper grammar is for properly
defining tail calls modulo cons.  The lazy way would be to
allow, in tail position, a single constructor application itself
containing a tail position, that is, add a case $\constr K {(\fam i
{e_i}, \hole, \fam j {e_j})}$ to the definition of $\Tcont$.  This
would capture most of the cases presented above.  However, such a lazy
implementation would be brittle: for instance, a partially unrolled
version of \code{map} below would not benefit from the optimization.

\begin{lstlisting}
let rec umap f xs =
  match xs with
  | [] -> []
  | [x] -> [f x]
  | x1 :: x2 :: xs ->
  	f x1 :: f x2 :: umap f xs
\end{lstlisting}

This would make the performance-seeking developer unhappy, as they
would have to choose between our optimization or the performance
benefits of unrolling.  To make them happy again, we at least need to
consider \textbf{repeated} constructor applications.  Using the
grammar $\Ccont$ from Figure~\ref{fig:consgram}, we could consider all
the decompositions $\plug \Tcont \Ccont$, and extract the inner hole
from the nested $\Ccont$ context.


However, this approach would still be somewhat fragile, and some very
reasonable program transformations on the nested TMC call would break
it.  For instance, it is not possible to locally let-bind parts of the
function application, or to perform a match ultimately leading to
a TMC call inside a constructor application. In our new grammar $U$,
instead of adding a case
$\constr K {(\fam i {e_i}, \hole, \fam j {e_j})}$ that forces
constructors to occur only at the "leaves" of a context, we add a case
$\constr K {(\fam i {e_i}, U, \fam j {e_j})}$, allowing arbitrary
interleavings of constructors and tail-preserving contexts. This gives
a more natural and less surprising definition of the tail positions
modulo cons. This grammar is depicted in Figure~\ref{fig:tmcgram}.

\begin{mathparfig}{fig:tmcgram}{Tail modulo cons contexts}
  \begin{array}{r@{~}r@{~}l@{\quad}l}
    \Set{TailModConsCtx} \ni \TMCcont
    & \bnfeq
    &
    \\
    & \bnfor
    & \hole
    &
    \\
    & \bnfor
    & \seq e \TMCcont
    &
    \\
    & \bnfor
    & \letin x e \TMCcont
    &
    \\
    & \bnfor
    & \match e {\fam j {\clause {p_j} \TMCcont }}
    &
    \\
    & \bnfor
    & \constr K {(\fam i {e_i}, \TMCcont, \fam j {e_j})}
  \end{array}
\end{mathparfig}

If an expression $e_0$ is of the form
$\plug \TMCcont {\fam i {e_i}, \app f e, \fam j {e_j}}$, we say that
the plugged subterms are in tail position \emph{modulo constructor} in $e_0$,
and in particular $\app f e$ is a tail call \emph{modulo constructor}. We
also define tail positions \emph{strictly} modulo cons as the
tail positions modulo cons which are not regular tail
positions.

Notice that there is a subtlety here: the term
$\constr K {(\app{f_1}{e_1}, \app{f_2}{e_2})}$ admits two
\emph{distinct} context decompositions, one with
$U_1 \defeq \constr K {(\hole, \app{f_2}{e_2})}$ where $\app{f_1}{e_1}$
is a tail call modulo cons, the other with
$U_2 \defeq \constr K {(\app{f_1}{e_1}, \hole)}$ where $\app{f_2}{e_2}$
is a tail call modulo cons. (This is intentional, obtained by
allowing a single sub-context in the constructor rule of $\TMCcont$.)
We can transform this term such that either one of the calls become
tail-calls, but not both. In other words, the notion of being ``in
tail position modulo cons'' depends on the decomposition context
$\TMCcont$.

Our implementation has to decide which context decomposition to
perform. It does not make choices on the user's behalf: in such
ambiguous situations, it will ask the user to disambiguate by adding
a \code{[@tailcall]} attribute to the one call that should be made
a tail-call.

Remark: our grammar for $U$ is \emph{maximal} in the sense that in
each possible context decomposition of a term, all tail positions
modulo cons are inside a hole of the $U$ context. It would be possible
to abandon maximality by allowing arbitrary terms $e$
(containing no hole) as a context. We avoided doing this, as it would
introduce ambiguities in the grammar of context (the program $(a; b)$
can be parsed using this $e$ case directly, or using the $e; U$
rule first), so that operations defined on contexts would depend on
the parse tree of the context in the grammar.

\section{TRMC functions of interest}
\label{sec:examples}

Many functions that consume and produce lists are
tail-recursive-modulo-cons, in the sense that all they have a TMC
decomposition where all recursive calls are in TMC position. Notable
functions include \code{map}, as already discussed, but also for
example:

\begin{lstlisting}
let[@tail_mod_cons] rec filter p = function
| [] -> []
| x :: xs -> if p x then x :: filter p xs else filter p xs

let[@tail_mod_cons] rec merge cmp l1 l2 =
  match l1, l2 with
  | [], l | l, [] -> l
  | h1 :: t1, h2 :: t2 ->
      if cmp h1 h2 <= 0
      then h1 :: merge cmp t1 l2
      else h2 :: merge cmp l1 t2
\end{lstlisting}

TMC is not useful only for lists or other ``linear'' data types, with
at most one recursive occurrence of the datatype in each
constructor.

\paragraph{A non-example} Consider a \code{map} function on binary
trees:
\begin{lstlisting}
let[@tail_mod_cons] rec map f = function
| Leaf v -> Leaf (f v)
| Node(t1, t2) -> Node(map f t1, (map[@tailcall]) f t2)
\end{lstlisting}
In this function, there are two recursive calls, but only one of them
can be optimized; we used the \code{[@tailcall]} attribute to direct
our implementation to optimize the call to the right child. This is
actually a \emph{bad} example of TMC usage in most cases, given that
\begin{itemize}
\item If the tree is arbitrary, there is no reason that it would be
  right-leaning rather than left-leaning. Making only the right-child
  calls tail-calls does not protect us from stack overflows.
\item If the tree is known to be balanced, then in practice the depth
  is probably very small in both directions, so the TMC transformation
  is not necessary to have a well-behaved function.
\end{itemize}

\paragraph{Yes-examples from our real world} There \emph{are} interesting
examples of TMC-transformation on functions operating on tree-like
data structures, when there are natural assumptions about which child
is likely to contain a deep subtree. The OCaml compiler itself
contains a number of them; consider for example the following function
from the \code{Cmm} module, one of its lower-level program
representations:
\begin{lstlisting}
let[@tail_mod_cons] rec map_tail f = function
  | Clet(id, exp, body) ->
      Clet(id, exp, map_tail f body)
  | Cifthenelse(cond, ifso, ifnot) ->
      Cifthenelse(cond, map_tail f ifso, (map_tail[@tailcall]) f ifnot)
  | Csequence(e1, e2) ->
      Csequence(e1, map_tail f e2)
  | Cswitch(e, tbl, el) ->
      Cswitch(e, tbl, Array.map (map_tail f) el)
  [...]
  | Cexit _ | Cop (Craise _, _, _) as cmm ->
      cmm
  | Cconst_int _ | Cvar _ | Ctuple _ | Cop _ as c ->
      f c
\end{lstlisting}

This function is traversing the ``tail'' context of an arbitrary
program term -- a meta-example! The \code{Cifthenelse} node acts as
our binary-node constructor, we do not know which side is likely to be
larger, so TMC is not so interesting. The recursive calls for
\code{Cswitch} are not in TMC position. But on the other hand the
\code{Clet}, \code{Csequence} cases are very beneficial to have in
TMC: while they have several recursive subtrees, they are in practice
only deeply nested in the direction that is turned into a tailcall by
the transformation. The OCaml compiler does sometimes encounter
machine-generated programs with a unusually long sequence of either
constructions, and the TMC transformation may very well avoid a stack
overflow in this case.

Another example would be
\href{https://github.com/ocaml/ocaml/pull/9636}{\#9636}, a patch to
the OCaml compiler proposed by Mark Shinwell, to get
a partially-tail-recursive implementation of the ``Common
Subexpression Elimination'' (CSE) pass. Xavier Leroy remarked that the
existing implementation in fact fits the TMC fragment. Not all
recursive calls become tail-calls (this would require a more powerful
transformation or a longer, less readable patch), but the behavior
of TMC on the unchanged code matches the tail-call-ness proposed in
the human-written patch.



\section{Modular TMC transformation}

A good way of thinking about our TMC transformation is as follows.  We
want to transform a tail context modulo cons $\TMCcont$ into
a regular tail context $\Tcont$, where tail calls modulo cons
have been replaced by regular tail calls, but to the DPS version of
the callee.  More precisely, given a term $e$, we will build its DPS
version as follows:
\begin{itemize}
	\item First, we find a decomposition of $e$ as
		$\plug \TMCcont {e_1, \dots, e_n}$
		which identifies the tail positions modulo cons.  We want
		this decomposition to capture as much of the TMC calls to
		DPS-enabled functions (functions marked with the
		\code{[@tail_mod_cons]} attribute) as possible.

	\item Once the decomposition is selected, we transform the context
		$\setref x y \TMCcont$ (where $x$ and $y$ are fresh variables not
		appearing in $\TMCcont$) into a context
		$\plug \Tcont {\setref {x_1} {y_1} \hole, \dots, \setref {x_n}
		{y_n} \hole}$.  This transformation is effectively moving the
		assignment from the root to the leaves of the context $\TMCcont$.

	\item Finally, we replace the assignments $\setref x y
		{\app f e}$ by calls $\app {\dps f} {( (x, y), e)}$
		when the callee $f$ has a DPS version $\dps f$,
		introducing calls in tail position.
\end{itemize}

However, this transformation is not enough: we also need to transform
the code of the original function to call $\dps f$ in the recursive
case. There is a naive way to do it, which is suboptimal, as we
explain, before showing how we do it, in
\myfullref{Section}{subsect:transform-direct}.

Finally, \myfullref{Section}{subsec:constructor-compression}
highlights an optimization made by our implementation which generates
cleaner code for TMC calls inside nested constructor applications.




\subsection{The DPS transformation}
\label{subsec:transform-dps}

We now present in detail the transformation used to build the body of
the transformed $\dps f$ function from a function definition $\letrec
f x e$.  As a reminder, the semantics of $\app {\dps f} {( (d, i),
{e'})}$ should be the same as $\setref d i {\app f {e'}}$, and the
body of $\dps f$ should replace tail calls modulo cons in $e$
with regular tail calls to the DPS versions of the callee.  We only
present the transformation for unary functions: the general case
follows by using tuple for n-ary functions.  Our implementation
handles fully applied functions of arbitrary arity, treating them
similarly as the equivalent unary function with a tuple argument.

We first find a decomposition of $e$ as $\plug \TMCcont {e_1, \dots,
e_n}$.  Recall that the TMC transformation depends on this
decomposition, as we have multiple choices for the decomposition of a
constructor $\constr K {(e_1, \dots, e_n)}$.  We use the following
logic:

\begin{itemize}
	\item If none of the $e_i$ contains calls in TMC position to a
		function which has a DPS version (a \emph{TMC candidate}), use the
		decomposition $e = \plug \hole e$.
	\item If exactly one of the $e_i$ contains such a TMC candidate, or
		if exactly one of the $e_i$ contains a TMC candidate marked with
		the \code{[@tailcall]} annotation, name it $e_j$ and use the
		decomposition $e = \plug {(\constr K {(e_1, \dots, e_{j - 1},
		\hole, e_{j + 1}, \dots, e_n)}} {e_j}$.
	\item Otherwise, report an error to the user, indicating the
		ambiguity, and requesting that one (or more) \code{[@tailcall]}
		annotations be added.
\end{itemize}

For other constructs, we only decompose them if at least one of their
component has a TMC candidate; for instance, $\letin x e {e'}$ gets
decomposed into $\plug \hole {\letin x e {e'}}$ unless $e'$ contains
TMC candidates.  This avoids needlessly duplicating assignments.

Once we obtained the decomposition as a tail context modulo
constructors, we transform said context into a tail context where each
expression in tail position is an assignment.  We write $\dpred {d}
{n} {\TMCcont} {\plugfam \Tcont i {\setref {d_i} {n_i} \hole}}$
to signify that the context ${\plugfam \Tcont i {\setref {d_i}
{n_i} \hole}}$ is obtained by performing the DPS transformation on
the TMC context $\TMCcont$ with destination $d . n$. The rules
describing this transformation are shown below.
\begin{mathpar}
  \infer{ }
	{\dpred d n \hole {\plug \hole {\setref d n \hole}}}

  \infer
  {\dpred d n \TMCcont
    {\plugfam \Tcont i {\setref {d_i} {n_i} \hole}}}
  {
    \dpred d n {\letin x e \TMCcont}
      {\letin x e {\plugfam \Tcont i {\setref {d_i} {n_i} \hole}}}
  }

  \infer
  {\forall j,\quad
    \dpred d n {\TMCcont_j}
      {\plugfam {T_j} {i_j} {\setref {d_{i_j}} {n_{i_j}} \hole}}}
  {\dpred d n
     {\match e {\fam j {\clause {p_j} \TMCcont_j}}}
     {\match e {\fam j {\clause {p_j} {\plugfam {T_j} {i_j} {\setref {d_{i_j}} {n_{i_j}} \hole}}}}}
  }

  \infer
  {n' = \vert I \vert + 1
   \\
   \dpred {d'} {n'} \TMCcont
     {\plugfam \Tcont l {\setref {d'_l} {n'_l} \hole}}
  }
  {
    \dpred d n
      {\constr K {(\fam {i \in I} {e_i}, \TMCcont, \fam j {e_j})}}
      {\begin{array}{l}
         \letin {d'} {\constr K {(\fam {i \in I} {e_i}, \kwd{Hole}, \fam j {e_j})}} {}
         \\
         \quad \seq {\setref d n {d'}} {}
         \\
         \quad \plugfam \Tcont l {\setref {d'_l} {n'_l} \hole}
       \end{array}}
    }
\end{mathpar}

Most cases are straightforward: for constructs whose tail context
modulo cons is also a regular tail context, we simply apply the
transformation into said tail context.  The two important cases are
the one of a hole, where we introduce an assignment to the result of
evaluating the hole, and the case of a constructor, where we ``reify''
the hole by using a placeholder value, fill the current destination,
and recursively transform the TMC context with the newly created hole
as a new destination $d'.n'$.\footnote{It may look like $d'$ is only
  used once in the right-hand-side of the conclusion of this rule, so
  its binding could be inlined, but this $d'$ is used in the last
  premise and will occur in $U$.}

After this transformation, we can now build the term
$\plugfam \Tcont i {\setref {d_i} {n_i} {e_i}}$ where the $e_i$ are
the subterms in the initial decomposition $\plugfam \TMCcont i {e_i}$
of the body of our recursive function.  Finally, for each $e_i$ with shape
$\app {f_i} {e'_i}$ where $f_i$ has a DPS version $f_i^{dps}$, we can
replace $\setref {d_i} {n_i} {e_i}$ with
$\app {{f_i}^{dps}} {((d_i, n_i), e_i)}$, yielding the final result of
the DPS transformation.

We remark again that the \emph{only} calls in tail position in the
transformed term are the assignments we have just transformed into
calls to a DPS version of the original callee.  Indeed, any other tail
call in the initial decomposition has been replaced by an
assignment. We "lose" a tail call when we go from the "transformed"
world back to the "direct-style" world -- a CPS transformation would
work similarly, transforming $\app f e$ into $\app k (\app f e)$ if
$f$ has no CPS version.

\subsection{The ``direct'' transformation}
\label{subsect:transform-direct}

As we just noted, tail calls in the original function are tail
calls in the DPS version only if their callee also have a DPS version (e.g.
in the common case of a recursive function).  Tail calls where the
callee didn't have a DPS version are no longer in tail position.  As
such, the simple and lazy way to call into $\dps f$ in the body of
$f$, namely, introducing a destination and calling $\dps f$, is
suboptimal, as it could make previously-tail recursive paths in $f$ no
longer tail recursive, and the programmer may rely on those being tail
recursive.  Instead, we will ensure that calls to $\dps f$ only
happen inside a constructor application: this way, all the
pre-existing tail calls will be left untouched.

The transformation is very similar as the DPS transformation (in fact,
all of the ``boring'' cases are identical), and we will reuse the same
context decomposition of $e$ as $\plug \TMCcont {e_1, \dots, e_n}$.
We again perform a context rewriting on $\TMCcont$, but now the output
is an arbitrary context $E$.

\begin{mathpar}
	\inferrule{ }{\hole \dired \hole}

	\inferrule{\TMCcont \dired E}{
		{\letin x e \TMCcont} \dired {(\letin x e E)}
	}

	\inferrule{\forall j,\; \TMCcont_j \dired E_j}{
		{\match e {\fam j {\clause {p_j} \TMCcont_j}}}
			\dired {(\match e {\fam j {\clause {p_j} E_j}})}
	}

	\inferrule
        {n = \vert I \vert + 1
         \\
         \dpred d n \TMCcont {\plugfam \Tcont l {\setref {d_l} {n_l} \hole}}}
         {
		{\constr K {(\fam {i \in I} {e_i}, \TMCcont, \fam j {e_j})}}
			\dired
			{\begin{array}{l}
                           \letin {d} {\constr K {(\fam {i \in I} {e_i}, \kwd{Hole}, \fam j {e_j})}} {}
                           \\
                           \quad \seq {\plugfam \Tcont l {\setref {d_l} {n_l} \hole}} {}
                           \\
                           \quad d
			\end{array}}
		}
\end{mathpar}

This transformation leaves the regular tail positions unchanged, but
switches to the DPS version for tail positions strictly modulo cons.
We then replace again tail assignments of a call to a tail call to the
DPS version of the callee.  Note that this time, calls to the DPS
version are not in tail position (we need to return the computed
value): we simply introduce a fresh destination so that we can call
into the DPS version.

The presentation by \cite*{minamide} suggests a slightly different
encoding, where we would pass a third extra argument to the DPS
version: the location of the final value to be returned.  This would
allow tail calls into the DPS version, at the cost of an extra
argument.  This may look compelling, but in practice not so much,
because calls from the DPS version back into the ``direct'' world will
never be in tail position, and we only end up paying a constant factor
more frames.

\subsection{Compression of nested constructors}
\label{subsec:constructor-compression}

Consider a function such as the partially unrolled \code{map} shown
above.   It has two nested constructor applications, and the DPS
transformation as described above will generate the code on the left
below for the \code{umap_dps} version.  This is unsatisfactory, as it
introduces needless writes that the OCaml compiler does not eliminate.
Instead, we would want to generate the nicer code on the right.

\hspace{-1.6em}
\begin{minipage}{0.5\linewidth}
\begin{lstlisting}
let rec umap_dps dst i f = function
| [] ->
	dst.i <- []
| [x] ->
	dst.i <- [f x]
| x1 :: x2 :: xs ->
	let dst1 = f x1 :: Hole in
	dst.i <- dst1;
	let dst2 = f x2 :: Hole in
	dst1.1 <- dst2;
	umap_dps dst2 1 f xs
\end{lstlisting}
\end{minipage}
\hfill
\begin{minipage}{0.5\linewidth}
\begin{lstlisting}
let rec umap_dps dst i f = function
| [] ->
	dst.i <- []
| [x] ->
	dst.i <- [f x]
| x1 :: x2 :: xs ->
	let y1 = f x1 in
	let y2 = f x2 in
	let dst' = y2 :: Hole in
	dst.i <- y1 :: dst';
	umap_dps dst' 1 f xs
\end{lstlisting}
\end{minipage}

Notice that in the nicer code, we need to let-bind constructor
arguments to preserve execution order.  We implement this optimization
by keeping track, in the rules for the $dps$ transformation, of an additional
``constructor context''.  Before, we were conceptually preserving
the semantics of $\setref d n \TMCcont$; now, we will be preserving
the semantics of $\setref d n {\plug \Ccont \TMCcont}$ for some
constructor context $\Ccont$.  $\Ccont$ represents delayed constructor
applications, which we will perform later --- typically, immediately
before calling a DPS-transformed function.

The new rules, written
$\dpredo d n \Ccont \TMCcont {\plugfam \Tcont i {\setref {d_i} {n_i} {\Ccont_i}}}$,
are shown in \myref{Figure}{fig:dps-with-compression}.  The rules for
$\kwd{let}$ and $\kwd{match}$ are unchanged, except that we pass the
unchanged constructor context $\Ccont$ recursively.

\begin{mathparfig}{fig:dps-with-compression}{DPS transformation, with constructor compression}
	\inferrule[DPS-Hole-Opt]{ }{\dpredo d n \Ccont \hole {\plug \hole {\setref d n
	{\Ccont}}}}

	\inferrule[DPS-Reify]
        {n' = \vert I \vert + 1
         \\
         \dpredo {d'} {n'} \hole \TMCcont {\plugfam \Tcont l {\setref {d_l} {n_l} {\Ccont_l}}}}
        {
          \dpredo d n
            {\plug \Ccont {\constr K {(\fam i {e_i}, \hole, \fam j {e_j})}}} \TMCcont
            {\begin{array}{l}
              \letin {d'} {\constr K {(\fam {i \in I} {e_i}, \kwd{Hole}, \fam j {e_j})}} {} \\
              \quad \seq {\setref d n {\plug \Ccont {d'}}} {}
              \\
              \quad {\plugfam \Tcont l {\setref {d_l} {n_l} {\Ccont_l}}}
            \end{array}}
	}

	\inferrule[DPS-Constr-Opt]
        {n' = \vert I \vert + 1
         \\
         \dpredo {d'} {n'} {\plug \Ccont {\constr K (\fam {i \in I} v_i, \hole, \fam j {e_j})}} \TMCcont
         {\plug \Tcont {\fam l {\setref {d_l} {n_l} {\Ccont_l}}}}}
        {
          \dpredo d n \Ccont {\constr K {(\fam {i \in I} {e_i}, \TMCcont, \fam j {e_j})}}
            {\begin{array}{l}
              \letfamin {i \in I} {v_i} {e_i} {} \\
              \quad \letfamin j {v_j} {e_j} {} \\
              \quad \plugfam \Tcont l {\setref {d_l} {n_l} {\Ccont_l}}
            \end{array}}
        }

	\inferrule{\dpredo d n \hole \TMCcont {\plug \hole
	{\setref d i {\Ccont_i}}}}{\dpred d n \TMCcont {\plug
	\hole {\setref d i {\Ccont_i}}}}
\end{mathparfig}

The constructor rule is now split in two parts: the
\textsc{DPS-Constr-Opt} rule adds the new constructor to the delayed
constructor context, and the \textsc{DPS-Reify} rule generates an
assignment for the constructor context.  Note that the rules for
$\rightsquigarrow_{dps}$ are no longer deterministic: the
\textsc{DPS-Reify} can apply whenever the delayed constructor stack is
nonempty.
%
%
We perform the reification in two cases.  The first one is before
generating a call to a DPS-transformed version, because we need
a concrete destination for that.  The second one is that we keep track
of whether a subterm would duplicate the delayed context (e.g. due to
a $\kwd{match}$) and immediately apply the reification after
a constructor in that case.

Notice that a similar optimization could be made in the $direct$
transformation (in fact, our implementation does just that):  the goal
of switching to the $dps$ mode in a constructor context is simply to
provide a destination to the inner TMC calls.  We can, without loss of
generality, only switch to $dps$ when there is a TMC call in tail
position in the recursive argument (i.e. there will be no
opportunities to introduce a destination in a subterm).

Finally, we note that in our implementation we perform all of the
transformations (dps, direct, as well as the computation of the
auxiliary information such as whether there are TMC opportunities in
the term and whether we need to provide a destination to benefit from
a switch to the DPS mode) in a single pass over the terms.

\section{Design issues}

\subsection{Non-issue: Flambda and Multicore}

Some readers may wonder whether introducing mutation to build
immutable data structures could be an issue with other subsystems of
the OCaml implementation that perform fine-grained mutability
reasoning, notably the Flambda optimizer and the Multicore runtime.

The answer is that there is no issue, move along! The OCaml value
model (even under Multicore) already contains the notions that
(immutable) values may start ``uninitialized'' and eventually be
filled by ``initializing'' writes -- this is how immutable values are
constructed from the C FFI, for example. In 2016, in preparation for
our TMC work, Frédéric Bour extended the Lambda intermediate
representation with an explicit notion of ``initializing'' write
(\href{https://github.com/ocaml/ocaml/pull/673}{\#673}), so that the
information is explicit in the generated code, and thus perfectly
understood by Flambda.

\subsection{Non-linear control operators}

The TMC transformation makes the assumption that partially-initialized
values (with a \code{Hole} placeholder) have a unique owner, and will
be initialized into complete values by a single write. This assumption
can be violated by the addition of control operators to the language,
such as \code{call/cc} or \code{delim/cc}. The problem comes from the
\emph{non-linear} usage of continuations, where the same continuation
is invoked twice.

In practice this means that we cannot combine the external
(but magical)
\href{http://okmij.org/ftp/continuations/implementations.html#caml-shift}{delim/cc}
library with our TMC transformation. Currently the only solution is to
disable the TMC transformation for delimited-continuation users
(at the risk of stack overflows), but in the future we could perform
a more general stack-avoiding transformation, such as
a continuation-passing-style transformation, for \code{delim/cc}
users.

(We considered intermediate approaches, based on detecting
multi-writes to a partially-initialized value, and copying the partial
value on the fly. This works for lists, where the position of the
hole is, but we did not manage to define this approach in the general
case of arbitrary constructors.)

\subsection{Evaluation order}

If a call inside a constructor is in TMC position, our transformation
ensures that it is evaluated ``last''. For example, in the program
$\constr K {(e_1, \app f {e_2}, e_3)}$, if $f$ is the TMC call, we
know that $e_1$ and $e_3$ will be evaluated (in some order) before
$\app f {e_2}$ in the transformed program.

In OCaml, the order of evaluation order of constructor arguments is
implementation-defined; the evaluation-order resulting
from the TMC transform is perfectly valid for the source
program. However, in this case it is also \emph{different} from the
one you would typically observe on the unmodified program -- most
implementations use either left-to-right or right-to-left.

We consider that it is reasonable that an \emph{explicit}
transformation would change the evaluation order -- especially as it
remains a valid order for the source program. Reviewers have found
this to be an issue, and suggested instead to forbid having
potentially-side-effectful arguments in TMC constructor applications:
in this example, if we restrict $e_1$ and $e_3$ to be values
(or variables), we cannot observe a difference anymore.

In several cases this forces the user to \code{let}-bind the arguments
beforehand, explicitly expressing an evaluation order. This is
a sensible design, but in our experience many functions that would
benefit from the TMC transformation are not written in this style, and
converting them to be in this style would be a bothersome and invasive
change, raising the barrier to entry of the \code{[@tail_mod_cons]}
annotation -- without much benefits in terms of evaluation-order, as
functions that need to enforce a specific evaluation order should use
explicit \code{let}-bindings anyway. This is for example the case of
the most interesting example of \myfullref{Section}{sec:examples}, the
\code{tail_map} function on a compiler intermediate representation.

\subsection{Stack usage guarantees}

This document presents a precisely-defined subset of functions that
can be TMC-transformed (they must be decomposable through a TMC context
$\TMCcont$, with a TMC-specialized function call in tail-position or,
preferably, strictly in tail-position modulo cons).

For many recursive functions in this subset, the TMC-transformation
returns a ``tail-recursive function'', using a constant amount of
stack space. However, many other functions are not ``tail-recursive
modulo cons'' (TRMC) in this sense. Initially we wanted to restrict
our transformation to reject non-TRMC functions: the success of the
transformation would guarantee that the resulting function never
stack-overflows.

However, we realized that many functions we want to TMC-transform are
not in the TRMC fragment. The interesting \code{tail_map} function
from \myfullref{Section}{sec:examples} does not provide this
guarantee -- for instance, its stack usage grows linearly with the
nesting of \code{Cifthenelse} constructors in the \code{then}
direction.

\subsection{Transformation scope}

If a given function $f$ is marked for TMC transformation, what is the
scope of the code in which TMC calls to $f$ should be transformed? If
one tried to give a maximal scope, an answer could be: all calls to
$f$ in the program. This requires including information on which
functions have a DPS version in module boundaries, and thus in module
types (to enable separate compilation). We tried to find a smaller
scope that is easy to define, and would not require cross-module
information.

The second idea was the following ``minimal'' scope: when we have
a group of mutually-recursive functions marked for TMC, we should only
rewrite calls to those functions in the recursive bodies themselves,
not in the rest of the program. In
$\letrecfamin i {f_i} x {e_i} {e'}$, the calls to the $f_i$ in some
$e_j$ would get rewritten, but not in $e'$. This restriction makes
sense, given that generally the stack-consuming calls are done within
the recursive bodies, with only a constant number of calls in $e'$
that do not contribute to stack exhaustion.

However, consider the \code{List.flatten} function, which flattens
lists of lists into simple lists.

\begin{lstlisting}
let rec flatten = function
| [] -> []
| xs :: xss -> xs @ flatten xs
\end{lstlisting}

This function is not in the TRMC fragment. However, it can be
rewritten to be TMC-transformable by performing a local inlining of
the \code{@} operator to flatten two lists:

\begin{lstlisting}
let[@tail_mod_cons] rec flatten = function
| [] -> []
| xs :: xss ->
  let[@tail_mod_cons] rec append_flatten xs xss =
    match xs with
    | [] -> flatten xss
    | xs :: xss -> xs :: append_flatten xs xss
  in append_flatten xs xss
\end{lstlisting}

This definition contains a toplevel recursive function and a local
recursive function, and both are in the TMC fragment. However, to get
constant-stack usage it is essential that the call to
\code{append_flatten} that is \emph{outside} its recursive definition
be TMC-specialized. Otherwise it is not a tail-call anymore in the
transformed program.

For now we have decided to extend the ``minimal'' scope as follows:
for a recursive definition $\letrecfamin i {f_i} x {e_i} {e'}$, TMC
calls to the $f_i$ in the body $e'$ are not rewritten \emph{unless}
the whole term is itself in the body of a function marked for TMC. In
other words, the scope is ``minimal'' at the toplevel, but ``maximal''
within TMC-marked functions.

Another alternative is to stick to the minimal scope, and warn on the
\code{flatten} implementation above. It is still possible to write
a TMC \code{flatten} in the minimal scope, by extruding the local
definition into a mutual recursion:

\begin{lstlisting}
let[@tail_mod_cons] rec flatten = function
| [] -> []
| xs :: xss -> append_flatten xs xss

and[@tail_mod_cons] append_flatten xs xss =
  match xs with
  | [] -> flatten xss
  | xs :: xss -> xs :: append_flatten xs xss
\end{lstlisting}

Indeed, for local definitions it is always possible to rewrite the
term $\letrecfamin i {f_i} x {e_i} {e'}$ by moving $e'$ inside the
mutually-recursive part:
$\kwd{let}~\kwd{rec}~(\fam i {\app {f_i} x = e_i}, \app g \wild = e')~\kwd{in}~\app g {()}$. The question would be whether we want to force users to perform this transformation manually, and how to tell them that we expect it.

\subsection{Future work: Higher-order transformation}

We formalized the TMC-transformation on a first-order language, and
our implementation silently ignores the higher-order features of
OCaml. What would it mean to DPS-transform a higher-order function
such as $(\simplelet {\app {\app {\text{app}} f} x} {\app f x})$?

Our answer would be that the higher-order DPS-transformation takes
a pair of function $f$ and returns a pair
$(\direct f, \dps f)$, allowing to define
\begin{mathline}
\begin{array}{l@{~}l@{~}l@{~}l@{~}l@{~}l}
\kwd{let}~\text{app}^\mathsf{direct}
&
& (\direct f, \dps f)
& x
& =
& \app {\direct f} x
\\
\kwd{let}~\text{app}^\mathsf{dps}
& (d, i)
& (\direct f, \dps f)
& x
& =
& \app {\app {\dps f} {(d, i)}} x
\\
\end{array}
\end{mathline}

\subsection{Future work: Multi-destination TMC?}

The program below on the left, frustratingly, cannot be
TMC-transformed with our current definition or implementation. One can
manually express a \emph{multi-destination} DPS version, on the right,
but it is still unclear how to specify the fragment of input programs
that would be transformed in this way. (We discussed this with Pierre
Chambart.)

\hspace{-1.6em}
\begin{minipage}{0.45\linewidth}
\begin{lstlisting}
let rec partition p = function
| [] -> ([], [])
| x::xs ->
  let (yes, no) = partition p xs in
  if p x
  then (x :: yes, no)
  else (yes, x :: no)
\end{lstlisting}
\end{minipage}
\hfill
\begin{minipage}{0.65\linewidth}
\begin{lstlisting}
let rec partition_dps
  dst_yes i_yes dst_no i_no p xs
 = match xs with
| [] -> dst_yes.i_yes <- []; dst_no.i_no <- []
| x :: xs ->
  let dst_yes', i_yes', dst_no', i_no' =
    if p x then
      let dst' = x :: Hole in
      dst_yes.i_yes <- dst';
      dst', 1, dst_no, y_no
    else
      let dst' = x :: Hole in
      dst_no.i_no <- dst';
      dst_yes, i_yes, dst', 1
  in
  partition_dps
    dst_yes' i_yes' dst_right' i_right' p xs
\end{lstlisting}
\end{minipage}

\bibliography{tmc}

\appendix

\section{Performance evaluation}

In this section we present preliminary performance numbers for
the TMC-transformed version of \code{List.map}, which appear to
confirm the claim that this version is ``almost as fast as the naive,
non-tail-recursive version'' -- and supports lists of all length.

The performance results were produced by Anton Bachin's benchmarking
script \href{https://github.com/aantron/faster-map/}{faster-map},
which internally uses the
\href{https://github.com/janestreet/core_bench}{core-bench} library
that is careful to reduce measurement errors on short-running
functions, and also measures memory allocation. They are
``preliminary'' in that they were run on a personal laptop (running an
Intel Core i5-4500U -- Haswell), we have not reproduced results in
a controlled environment or on other architectures. We ran the
benchmarks several times, with variations, and the qualitative results
were very stable.

\subsection{The big picture}

We graph the performance ratio of various \code{List.map}
implementation, relative to the ``naive tail-recursive version'',
that accumulates the results in an auxiliary list that is reversed at
the end -- the first tail-recursive version of our Prologue. We
pointed out that this implementation is ``slow'' compared to the
non-tail-recursive version: for most input lengths it is the slowest
version, with other versions taking around 60-80\% of its runtime
(lower is better). One can also see that it is not \emph{that} slow,
it is at most twice as slow.

\newcommand{\bench}[1]{\textbf{#1}}

The other \code{List.map} versions in the graph are the following:
\begin{description}
\item[\bench{base}] The implementation of Jane Street's
  \href{https://github.com/janestreet/base}{Base} library
  (version 0.14.0). It is hand-optimized to compensate for the costs
  of being tail-recursive.
\item[\bench{batteries}] The implementation of the community-maintained
  \href{https://github.com/ocaml-batteries-team/batteries-included/}{Batteries}
  library. It is actually written in destination-passing-style, using
  an unsafe encoding with \code{Obj.magic} to unsafely cast a mutable
  record into a list cell. (The trick comes from the older Extlib
  library, and its implementation has a comment crediting Jacques
  Garrigue for the particular encoding used.)
\item[\bench{containers}] Is another standard-library extension by Simon
  Cruanes; it is the hand-optimized tail-recursive implementation we
  included in the Prologue.
\item[\bench{trmc}] is ``our'' version, the last version of the Prologue: the
  result of applying our implementation of the TMC transformation to
  the simple, non-tail-recursive version.
\item[\bench{stdlib}] is the non-tail-recursive version that is in the OCaml
  standard library. (All measurements used OCaml 4.10)
\item[\bench{stdlib unrolled 5x, trmc unrolled 4x}] are the result of manually
  unrolling the simple implementation (to go in the direction of the
  Base and Containers implementation); in the \bench{trmc} case, the
  result is then TRMC-transformed.
\end{description}

Our expectation, before running measurements, are that \bench{trmc}
should be about as fast as \bench{stdlib}, both slower than
manually-optimized implementations (they were tuned to compete with
the \bench{stdlib} version). We hoped that \bench{trmc unrolled 4x}
would be competitive with the manually-optimized
implementations. Finally, \bench{batteries} should be about on-par
with \bench{trmc}, as it is using the same implementation technique
(as a manual transformation rather than
a compiler-supported transformation).

\def\svgscale{0.8}
\graphicspath{{benchmarks/}}
\begingroup%
  \makeatletter%
  \providecommand\color[2][]{%
    \errmessage{(Inkscape) Color is used for the text in Inkscape, but the package 'color.sty' is not loaded}%
    \renewcommand\color[2][]{}%
  }%
  \providecommand\transparent[1]{%
    \errmessage{(Inkscape) Transparency is used (non-zero) for the text in Inkscape, but the package 'transparent.sty' is not loaded}%
    \renewcommand\transparent[1]{}%
  }%
  \providecommand\rotatebox[2]{#2}%
  \newcommand*\fsize{\dimexpr\f@size pt\relax}%
  \newcommand*\lineheight[1]{\fontsize{\fsize}{#1\fsize}\selectfont}%
  \ifx\svgwidth\undefined%
    \setlength{\unitlength}{450bp}%
    \ifx\svgscale\undefined%
      \relax%
    \else%
      \setlength{\unitlength}{\unitlength * \real{\svgscale}}%
    \fi%
  \else%
    \setlength{\unitlength}{\svgwidth}%
  \fi%
  \global\let\svgwidth\undefined%
  \global\let\svgscale\undefined%
  \makeatother%
  \begin{picture}(1,0.83333333)%
    \lineheight{1}%
    \setlength\tabcolsep{0pt}%
    \put(0,0){\includegraphics[width=\unitlength,page=1]{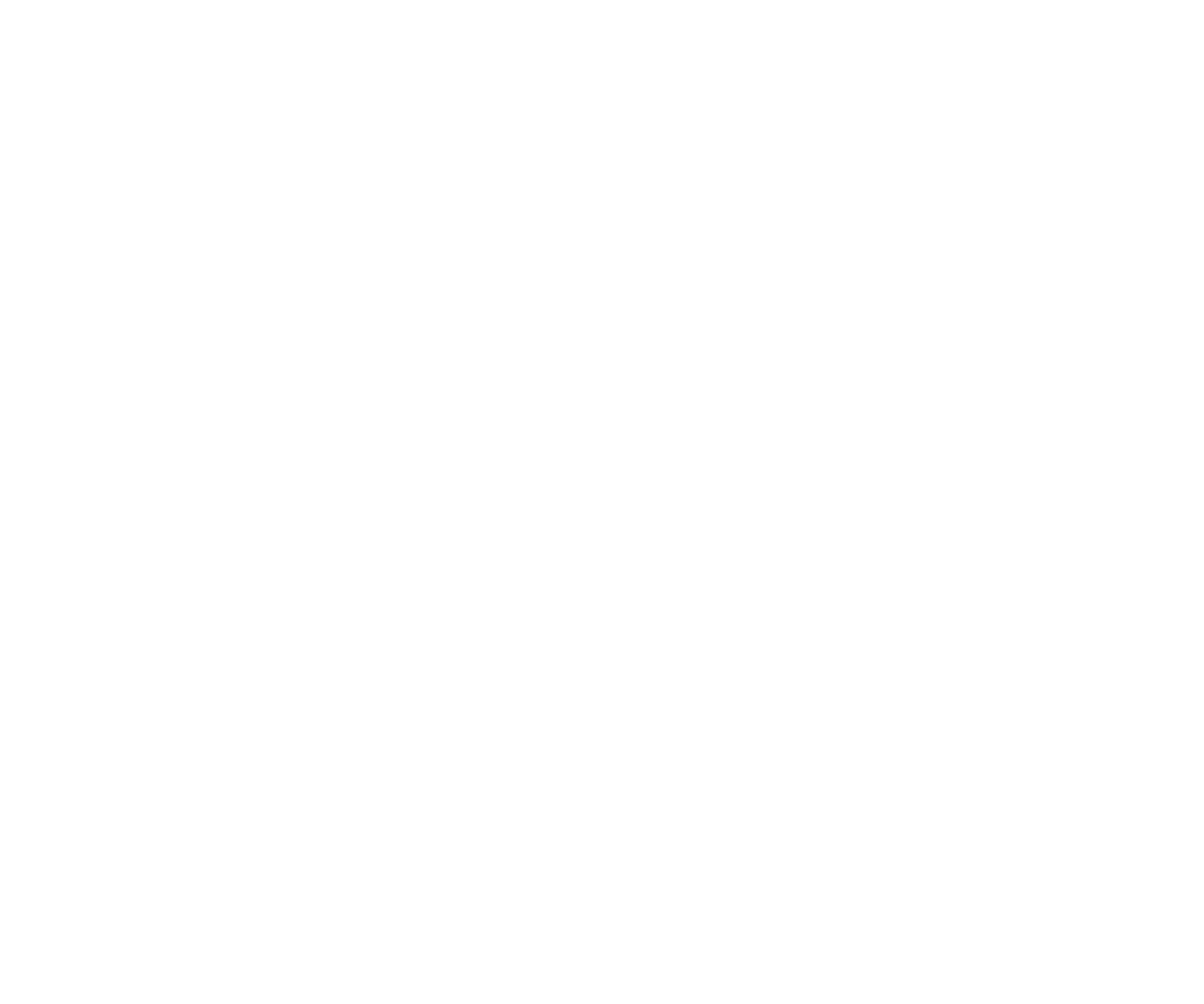}}%
    \put(0.12383333,0.10433333){\makebox(0,0)[rt]{\lineheight{1.25}\smash{\begin{tabular}[t]{r}0\end{tabular}}}}%
    \put(0,0){\includegraphics[width=\unitlength,page=2]{map_ignore.pdf}}%
    \put(0.12383333,0.21633333){\makebox(0,0)[rt]{\lineheight{1.25}\smash{\begin{tabular}[t]{r}20\end{tabular}}}}%
    \put(0,0){\includegraphics[width=\unitlength,page=3]{map_ignore.pdf}}%
    \put(0.12383333,0.32833333){\makebox(0,0)[rt]{\lineheight{1.25}\smash{\begin{tabular}[t]{r}40\end{tabular}}}}%
    \put(0,0){\includegraphics[width=\unitlength,page=4]{map_ignore.pdf}}%
    \put(0.12383333,0.4405){\makebox(0,0)[rt]{\lineheight{1.25}\smash{\begin{tabular}[t]{r}60\end{tabular}}}}%
    \put(0,0){\includegraphics[width=\unitlength,page=5]{map_ignore.pdf}}%
    \put(0.12383333,0.5525){\makebox(0,0)[rt]{\lineheight{1.25}\smash{\begin{tabular}[t]{r}80\end{tabular}}}}%
    \put(0,0){\includegraphics[width=\unitlength,page=6]{map_ignore.pdf}}%
    \put(0.12383333,0.6645){\makebox(0,0)[rt]{\lineheight{1.25}\smash{\begin{tabular}[t]{r}100\end{tabular}}}}%
    \put(0,0){\includegraphics[width=\unitlength,page=7]{map_ignore.pdf}}%
    \put(0.14,0.06933333){\makebox(0,0)[t]{\lineheight{1.25}\smash{\begin{tabular}[t]{c}1\end{tabular}}}}%
    \put(0,0){\includegraphics[width=\unitlength,page=8]{map_ignore.pdf}}%
    \put(0.27516667,0.06933333){\makebox(0,0)[t]{\lineheight{1.25}\smash{\begin{tabular}[t]{c}10\end{tabular}}}}%
    \put(0,0){\includegraphics[width=\unitlength,page=9]{map_ignore.pdf}}%
    \put(0.4105,0.06933333){\makebox(0,0)[t]{\lineheight{1.25}\smash{\begin{tabular}[t]{c}100\end{tabular}}}}%
    \put(0,0){\includegraphics[width=\unitlength,page=10]{map_ignore.pdf}}%
    \put(0.54566667,0.06933333){\makebox(0,0)[t]{\lineheight{1.25}\smash{\begin{tabular}[t]{c}1000\end{tabular}}}}%
    \put(0,0){\includegraphics[width=\unitlength,page=11]{map_ignore.pdf}}%
    \put(0.68083333,0.06933333){\makebox(0,0)[t]{\lineheight{1.25}\smash{\begin{tabular}[t]{c}10000\end{tabular}}}}%
    \put(0,0){\includegraphics[width=\unitlength,page=12]{map_ignore.pdf}}%
    \put(0.81616667,0.06933333){\makebox(0,0)[t]{\lineheight{1.25}\smash{\begin{tabular}[t]{c}100000\end{tabular}}}}%
    \put(0,0){\includegraphics[width=\unitlength,page=13]{map_ignore.pdf}}%
    \put(0.95133333,0.06933333){\makebox(0,0)[t]{\lineheight{1.25}\smash{\begin{tabular}[t]{c}$10^6$\end{tabular}}}}%
    \put(0.037,0.42){\rotatebox{90}{\makebox(0,0)[t]{\lineheight{1.25}\smash{\begin{tabular}[t]{c}Time relative to naive tail-recursive version (\%)\end{tabular}}}}}%
    \put(0.54566667,0.01683333){\makebox(0,0)[t]{\lineheight{1.25}\smash{\begin{tabular}[t]{c}List size (no. of elements)\end{tabular}}}}%
    \put(0.54566667,0.773){\makebox(0,0)[t]{\lineheight{1.25}\smash{\begin{tabular}[t]{c}Time elapsed (relative) – lower is better\end{tabular}}}}%
    \put(0.44516667,0.236){\makebox(0,0)[rt]{\lineheight{1.25}\smash{\begin{tabular}[t]{r}naive tail-rec.\end{tabular}}}}%
    \put(0,0){\includegraphics[width=\unitlength,page=14]{map_ignore.pdf}}%
    \put(0.44516667,0.2035){\makebox(0,0)[rt]{\lineheight{1.25}\smash{\begin{tabular}[t]{r}base\end{tabular}}}}%
    \put(0,0){\includegraphics[width=\unitlength,page=15]{map_ignore.pdf}}%
    \put(0.44516667,0.171){\makebox(0,0)[rt]{\lineheight{1.25}\smash{\begin{tabular}[t]{r}batteries\end{tabular}}}}%
    \put(0,0){\includegraphics[width=\unitlength,page=16]{map_ignore.pdf}}%
    \put(0.44516667,0.1385){\makebox(0,0)[rt]{\lineheight{1.25}\smash{\begin{tabular}[t]{r}containers\end{tabular}}}}%
    \put(0,0){\includegraphics[width=\unitlength,page=17]{map_ignore.pdf}}%
    \put(0.82666667,0.236){\makebox(0,0)[rt]{\lineheight{1.25}\smash{\begin{tabular}[t]{r}trmc\end{tabular}}}}%
    \put(0,0){\includegraphics[width=\unitlength,page=18]{map_ignore.pdf}}%
    \put(0.82666667,0.2035){\makebox(0,0)[rt]{\lineheight{1.25}\smash{\begin{tabular}[t]{r}trmc unrolled 4x\end{tabular}}}}%
    \put(0,0){\includegraphics[width=\unitlength,page=19]{map_ignore.pdf}}%
    \put(0.82666667,0.171){\makebox(0,0)[rt]{\lineheight{1.25}\smash{\begin{tabular}[t]{r}stdlib\end{tabular}}}}%
    \put(0,0){\includegraphics[width=\unitlength,page=20]{map_ignore.pdf}}%
    \put(0.82666667,0.1385){\makebox(0,0)[rt]{\lineheight{1.25}\smash{\begin{tabular}[t]{r}stdlib unrolled 5$\times$\end{tabular}}}}%
    \put(0,0){\includegraphics[width=\unitlength,page=21]{map_ignore.pdf}}%
  \end{picture}%
\endgroup%

\paragraph{Actual results}

In the first half of the graph, up to lists of size 1000, the results
are as we expected. There are three performance groups. The slow
tail-recursive baseline alone.  Then \bench{batteries}, \bench{stdlib}
and \bench{trmc} close to each other (\bench{batteries} does better
than the two other, which is surprising, possibly
a code-size effect). Then the versions using manual optimizations:
\bench{base}, \bench{containers}, and our \bench{unrolled} versions.

At the far end of the graph, with lists of size higher than $10^5$,
the results are interesting, and very positive: \bench{trmc} and
\bench{batteries} are the fastest versions, \bench{containers} is
slower. \bench{base} falls back to the slow tail-recursive version
after a certain input length, so its graph progressively joins the
baseline on larger lists. (Note: later versions of Base switched to
use an implementation closer to \bench{containers} in this regime.)

We also got performance results on \bench{stdlib} and \bench{stdlib
  unrolled} on large lists, by configuring the system to remove the
call-stack limit to avoid the stack overflows; their performance
profile is interesting and non-obvious, we discuss it in
\myfullref{Section}{subsec:ulimit}.

In the third quarter of the graph, for list sizes in the region
$[5.10^3;10^5]$, we observe surprising results where the
destination-passing-style version (\bench{trmc} and \bench{batteries})
become, momentarily, sensibly slower than the non-tail-recursive
\bench{stdlib} version. We discuss this strange effect in details in
\myfullref{Section}{subsec:promotion}, but the summary is that it is
mostly a GC effect due to the micro-benchmark setting (this strange
region disappears with a slightly different measurement), and would
not show up in typical programs.

\subsection{Promotion effects}
\label{subsec:promotion}

What happens in the $[5.10^3;10^5]$ region, where
destination-passing-style implementations seem to suddenly lose ground
against \bench{stdlib}? Promotion effects in the OCaml garbage
collector.

Consider a call to \code{List.map} on an input of length $N$, which is
in the process of building the result list. With the standard
\code{List.map}, the result is built ``from the end'': first the list
cell for the very last element of the list is allocated, then for the
one-before-last element, etc., until the whole list is created. With
the TMC-transformed \code{List.map}, the result list is built ``from
the beginning'': a list cell is first allocated for the first element
of the list (with a ``hole'' in tail positive) and written in the
destination, then a list cell for the second element is written in the
first cell's hole, etc., until the whole list is created.

The OCaml garbage collector is generational, with a small minor heap,
collected frequently, and a large major heap, collected
infrequently. When the minor heap becomes full, it is collected, and
all its objects that are still alive are promoted to the major heap.

What if the minor heap becomes full in the middle of the allocation of
the result list? Let's consider the average scenario where promotion
happens at cell $N/2$. In both cases (\bench{stdlib} or \bench{trmc}),
one half of the list cells are already allocated (the second or the
first half, respectively), and they get promoted to the major
heap. What differs is what happens \emph{next}. With the
non-tail-recursive implementation, the next list cell (in position
$N/2-1$) is allocated, pointing to the $(N/2)$-th cell, and the process
continues unchanged. In the destination-passing case, the next list
cell ($N/2+1$) is allocated, and it is written in the hole of the
$(N/2)$-th cell.

At each minor collection, the garbage collector (GC) needs to know
which objects are live in the minor heap. An object is live if it is
a root (a global value in the program, a value on the call stack,
etc.), if it is pointed to by a live object in the minor heap, or
a live object on the \emph{major} heap. The GC cannot afford to
traverse the large major heap to determine the latter category, so it
keeps track of all the pointers from the major to the minor heap; this
is called ``write barrier''. At this point in our new list's life,
writing the $(N/2+1)$-th cell to the $(N/2)$-th cell hits this write
barrier, and the $(N/2)$-th cell is added to the ``remembered set'', of
minor objects that are live due to the major heap. Trouble!

Hitting the write barrier has a small overhead, but this only happens
once in the middle of constructing a very large list. When the next
cell, the $(N/2+2)$-th list cell, gets written in the hole of the
$(N/2+1$)-th cell, both are in the minor heap, so the write barrier does
not come into play. Nothing happens until the next minor collection.

For the sake of simplicity, let's consider that the next minor
collection only happens after the whole result list has been
created. At this point, in the destination-passing-style version, the
$(N/2+1)$-th cell is in the remembered set, so it will be promoted by
the garbage collector to the major heap, along with all the objects
that it itself references; those are (at least) the list cells from
the middle to the end of the result list -- in total $N/2-1$ list
cells get promoted. In the \bench{stdlib} version, they corresponds to
the $N/2-1$ list cells at the beginning of the result list, allocated
after the previous minor collection. They will \emph{also} get
promoted... if the result list is still alive at this point!

To recapituate, the ``remainder'' of the result list is promoted in
all cases in the destination-passing version; it is only promoted in
the non-tail-recursive version if the result list is still alive.

``But this is silly'', you say, ``who would call \code{List.map} on
a very large list and drop the result immediately after?'' Well,
\href{https://github.com/aantron/faster-map/blob/5c1f2ca/tester/tester.ml#L57-L62}{this
  code}:
\begin{lstlisting}
let make_map_test name map_function argument : Core_bench.Bench.Test.t =
  Core_bench.Bench.Test.create ~name
    (fun () ->
      map_function ((+) 1) argument
      |> ignore)
\end{lstlisting}

\begin{remark}
  There \emph{are} real-world example where large results are
  short-lived. For example, consider a processing pipeline that calls
  \code{map}, and then \code{filter} on the result, etc.: the result
  of \code{map} may become dead quickly, if the lists are not too
  large. It is less likely to hit this promotion effect with
  medium-sized lists, but if this is all your application code is
  doing you will still see the effect once every $L/M$ calls, where
  $L$ is the length of your lists and $M$ the size of the minor heap,
  adding to a small but observable promotionoverhead.
\end{remark}

If we change the code to a version that guarantees that the result is
kept until at the next minor promotion, then there should be no
difference in promotion behavior. We did exactly this, and the new
graph looks like this:

\def\svgscale{0.8}
\graphicspath{{benchmarks/}}
\begingroup%
  \makeatletter%
  \providecommand\color[2][]{%
    \errmessage{(Inkscape) Color is used for the text in Inkscape, but the package 'color.sty' is not loaded}%
    \renewcommand\color[2][]{}%
  }%
  \providecommand\transparent[1]{%
    \errmessage{(Inkscape) Transparency is used (non-zero) for the text in Inkscape, but the package 'transparent.sty' is not loaded}%
    \renewcommand\transparent[1]{}%
  }%
  \providecommand\rotatebox[2]{#2}%
  \newcommand*\fsize{\dimexpr\f@size pt\relax}%
  \newcommand*\lineheight[1]{\fontsize{\fsize}{#1\fsize}\selectfont}%
  \ifx\svgwidth\undefined%
    \setlength{\unitlength}{450bp}%
    \ifx\svgscale\undefined%
      \relax%
    \else%
      \setlength{\unitlength}{\unitlength * \real{\svgscale}}%
    \fi%
  \else%
    \setlength{\unitlength}{\svgwidth}%
  \fi%
  \global\let\svgwidth\undefined%
  \global\let\svgscale\undefined%
  \makeatother%
  \begin{picture}(1,0.83333333)%
    \lineheight{1}%
    \setlength\tabcolsep{0pt}%
    \put(0,0){\includegraphics[width=\unitlength,page=1]{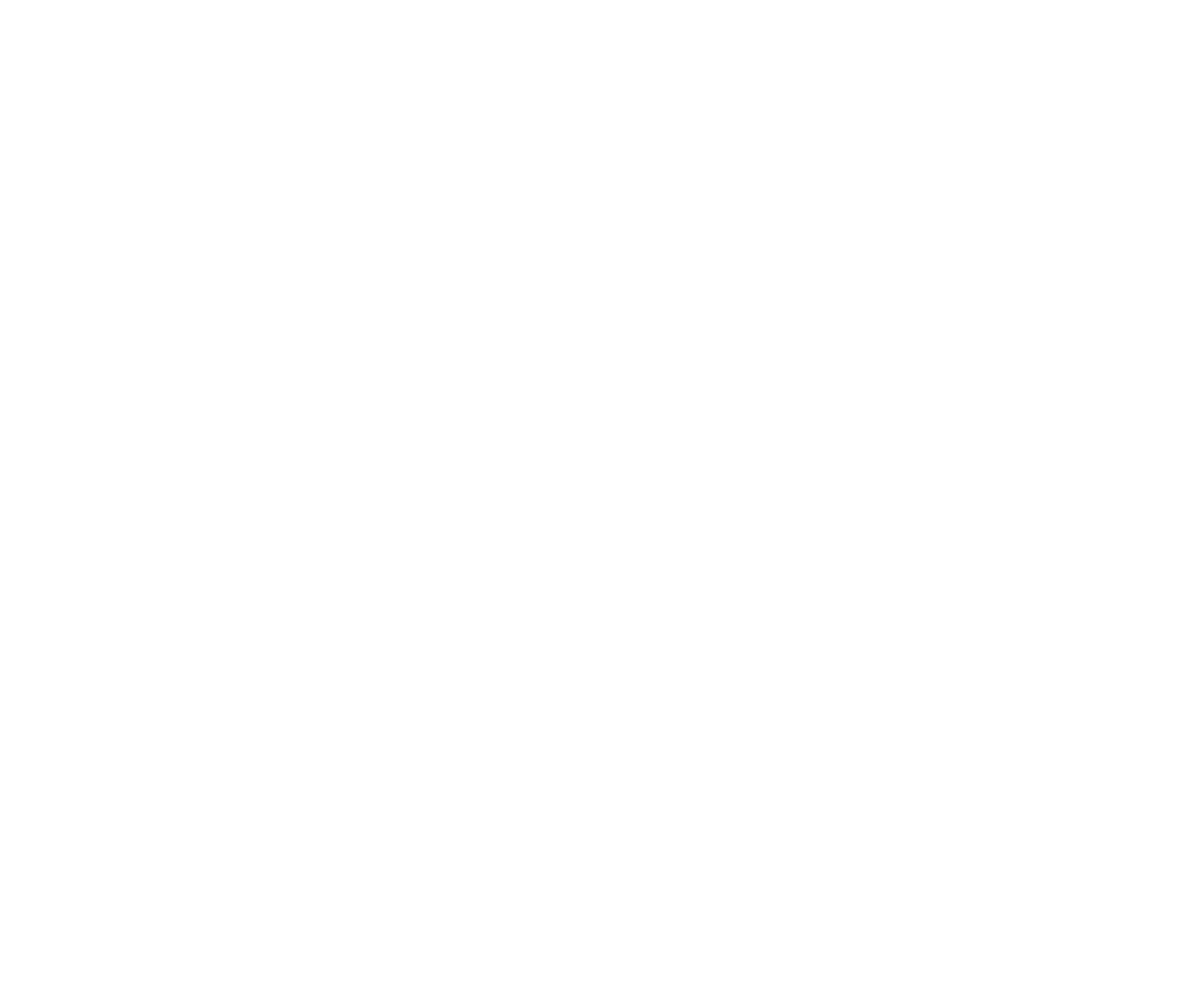}}%
    \put(0.12383333,0.10433333){\makebox(0,0)[rt]{\lineheight{1.25}\smash{\begin{tabular}[t]{r}0\end{tabular}}}}%
    \put(0,0){\includegraphics[width=\unitlength,page=2]{map_old.pdf}}%
    \put(0.12383333,0.21633333){\makebox(0,0)[rt]{\lineheight{1.25}\smash{\begin{tabular}[t]{r}20\end{tabular}}}}%
    \put(0,0){\includegraphics[width=\unitlength,page=3]{map_old.pdf}}%
    \put(0.12383333,0.32833333){\makebox(0,0)[rt]{\lineheight{1.25}\smash{\begin{tabular}[t]{r}40\end{tabular}}}}%
    \put(0,0){\includegraphics[width=\unitlength,page=4]{map_old.pdf}}%
    \put(0.12383333,0.4405){\makebox(0,0)[rt]{\lineheight{1.25}\smash{\begin{tabular}[t]{r}60\end{tabular}}}}%
    \put(0,0){\includegraphics[width=\unitlength,page=5]{map_old.pdf}}%
    \put(0.12383333,0.5525){\makebox(0,0)[rt]{\lineheight{1.25}\smash{\begin{tabular}[t]{r}80\end{tabular}}}}%
    \put(0,0){\includegraphics[width=\unitlength,page=6]{map_old.pdf}}%
    \put(0.12383333,0.6645){\makebox(0,0)[rt]{\lineheight{1.25}\smash{\begin{tabular}[t]{r}100\end{tabular}}}}%
    \put(0,0){\includegraphics[width=\unitlength,page=7]{map_old.pdf}}%
    \put(0.14,0.06933333){\makebox(0,0)[t]{\lineheight{1.25}\smash{\begin{tabular}[t]{c}1\end{tabular}}}}%
    \put(0,0){\includegraphics[width=\unitlength,page=8]{map_old.pdf}}%
    \put(0.27516667,0.06933333){\makebox(0,0)[t]{\lineheight{1.25}\smash{\begin{tabular}[t]{c}10\end{tabular}}}}%
    \put(0,0){\includegraphics[width=\unitlength,page=9]{map_old.pdf}}%
    \put(0.4105,0.06933333){\makebox(0,0)[t]{\lineheight{1.25}\smash{\begin{tabular}[t]{c}100\end{tabular}}}}%
    \put(0,0){\includegraphics[width=\unitlength,page=10]{map_old.pdf}}%
    \put(0.54566667,0.06933333){\makebox(0,0)[t]{\lineheight{1.25}\smash{\begin{tabular}[t]{c}1000\end{tabular}}}}%
    \put(0,0){\includegraphics[width=\unitlength,page=11]{map_old.pdf}}%
    \put(0.68083333,0.06933333){\makebox(0,0)[t]{\lineheight{1.25}\smash{\begin{tabular}[t]{c}10000\end{tabular}}}}%
    \put(0,0){\includegraphics[width=\unitlength,page=12]{map_old.pdf}}%
    \put(0.81616667,0.06933333){\makebox(0,0)[t]{\lineheight{1.25}\smash{\begin{tabular}[t]{c}100000\end{tabular}}}}%
    \put(0,0){\includegraphics[width=\unitlength,page=13]{map_old.pdf}}%
    \put(0.95133333,0.06933333){\makebox(0,0)[t]{\lineheight{1.25}\smash{\begin{tabular}[t]{c}$10^6$\end{tabular}}}}%
    \put(0.037,0.42){\rotatebox{90}{\makebox(0,0)[t]{\lineheight{1.25}\smash{\begin{tabular}[t]{c}Time relative to naive tail-recursive version (\%)\end{tabular}}}}}%
    \put(0.54566667,0.01683333){\makebox(0,0)[t]{\lineheight{1.25}\smash{\begin{tabular}[t]{c}List size (no. of elements)\end{tabular}}}}%
    \put(0.54566667,0.773){\makebox(0,0)[t]{\lineheight{1.25}\smash{\begin{tabular}[t]{c}Time elapsed (relative) – lower is better\end{tabular}}}}%
    \put(0.44516667,0.236){\makebox(0,0)[rt]{\lineheight{1.25}\smash{\begin{tabular}[t]{r}naive tail-rec.\end{tabular}}}}%
    \put(0,0){\includegraphics[width=\unitlength,page=14]{map_old.pdf}}%
    \put(0.44516667,0.2035){\makebox(0,0)[rt]{\lineheight{1.25}\smash{\begin{tabular}[t]{r}base\end{tabular}}}}%
    \put(0,0){\includegraphics[width=\unitlength,page=15]{map_old.pdf}}%
    \put(0.44516667,0.171){\makebox(0,0)[rt]{\lineheight{1.25}\smash{\begin{tabular}[t]{r}batteries\end{tabular}}}}%
    \put(0,0){\includegraphics[width=\unitlength,page=16]{map_old.pdf}}%
    \put(0.44516667,0.1385){\makebox(0,0)[rt]{\lineheight{1.25}\smash{\begin{tabular}[t]{r}containers\end{tabular}}}}%
    \put(0,0){\includegraphics[width=\unitlength,page=17]{map_old.pdf}}%
    \put(0.82666667,0.236){\makebox(0,0)[rt]{\lineheight{1.25}\smash{\begin{tabular}[t]{r}trmc\end{tabular}}}}%
    \put(0,0){\includegraphics[width=\unitlength,page=18]{map_old.pdf}}%
    \put(0.82666667,0.2035){\makebox(0,0)[rt]{\lineheight{1.25}\smash{\begin{tabular}[t]{r}trmc unrolled 4x\end{tabular}}}}%
    \put(0,0){\includegraphics[width=\unitlength,page=19]{map_old.pdf}}%
    \put(0.82666667,0.171){\makebox(0,0)[rt]{\lineheight{1.25}\smash{\begin{tabular}[t]{r}stdlib\end{tabular}}}}%
    \put(0,0){\includegraphics[width=\unitlength,page=20]{map_old.pdf}}%
    \put(0.82666667,0.1385){\makebox(0,0)[rt]{\lineheight{1.25}\smash{\begin{tabular}[t]{r}stdlib unrolled 5$\times$\end{tabular}}}}%
    \put(0,0){\includegraphics[width=\unitlength,page=21]{map_old.pdf}}%
  \end{picture}%
\endgroup%

This is the version that we consider representative of most
applciations, where data is long-lived enough that the subtle
promotion effects do not get into play. Notice in particular that, on
this version, \bench{trmc unrolled} is robustly better than all other
implementations. (\bench{stdlib unrolled} is still somewhat faster in
the previous ``awkward region'', and we are not sure why, nor do we
care very much.)

\subsection{ulimit}
\label{subsec:ulimit}

Using \code{ulimit -s unlimited}, we removed the call-stack limit on
our test machine, to test the speed of the non-tail-recursive
\code{List.map} on large lists. These behaviors tend to be unobserved
by OCaml users, which just get a program crash with
a \code{Stack_overflow}.

It is interesting that the \bench{stdlib} version gets progressively
slower on large inputs, until it matches the performance of the
tail-recursive baseline. Notice that the \bench{stdlib unrolled}
variant is also getting slower, although it starts with a good safety
margin.

Pierre Chambart suggested that this slowdown may result from stack
scanning: when the GC performs a minor collection, it scans the call
stack to find root pointers to minor objects. When the result list
gets much larger than the minor-heap size, many minor collections will
occur during the \code{List.map} call, with progressively larger call
stacks. The total overhead of the call-stack scanning is thus
quadratic; scanning the stack is \emph{fast}, but it eventually slows
down those implementations noticeably. (In theory it would only get
slower and slower as we increased the input size.)

Some implementations use the heap rather than a call stack;
either the tail-recursive \code{List.map} in OCaml, or
non-tail-recursive versions in a language that allocates call frames
on the heap. Notice that those implementations do not suffer from such
a quadratic slowdown, thanks to the generational GC: when the
corresponding ``heap frames'' are scanned, they get moved to the major
heap, and they will not get scanned again by minor collections,
avoiding the quadratic behavior at this level -- eventually enough
memory is consumed that the major heap will need to be traversed
regularly. It would be possible to change the OCaml runtime to use
a similar approach for the system call stack: the runtime could keep
track of which portions of the call stack have been traversed already,
to not scan them again on the next minor collection. In fact, it
seems that the OCaml runtime
\href{https://github.com/ocaml/ocaml/blob/41f0522/runtime/roots_nat.c#L306-L311}{implements}
such an optimization on
\href{https://github.com/ocaml/ocaml/blob/41f0522/runtime/caml/stack.h#L47-L49}{Power
  architectures} only (?!). In our case it is of little pratical
interest, however, speeding up scenarios that we never observe due to
the system stack limit.

\end{document}